\documentstyle[12pt]{article}
\title{Primordial Density Fluctuations in Phase Coupling Gravity}
\author{C. E. M.  Batista \thanks{e.mail: batista@ifi.unicamp.br} \\ Instituto de 
F\'{\i}sica Gleb
Wataghin\\  Universidade Estadual de Campinas\\ Cx. Postal 6165, SP 01183-090, Brazil \\ and \\ M.
Schiffer \thanks{e.mail: schiffer@ime.unicamp.br}\\ Instituto de Matem\'{a}tica Estat\'{\i}stica e
Ci\^{e}ncias da Computa\c{c}\~{a}o\\ Universidade Estadual de Campinas\\Cx. Postal 6065, SP
01183-090, Brazil} 
\date{}
\begin{document} 
\maketitle
\abstract{In this paper we study the evolution of primordial density perturbations in the
framework of Phase Coupling Gravity, proposed by Bekenstein \cite{bek}. We show that in the very 
early universe, these perturbations grow with an exponential-like behaviour} 
\newpage
\section{Introduction}
\subsection*{Dark matter: a must.} 
The issue of the material content of the Universe is one of the most actual and controversial 
problems in Modern Cosmology. Nucleosynthesis puts a very stringent bound on the amount of the barionic  
component $\Omega_B \approx 0.016 h^{-1}$ \cite{peebles}. On the other hand, the flatness problem
requires $\Omega = 1$, otherwise an extreme fine tuning of $\Omega(t)$ would be called for in the
early universe \cite{kolb}. For this an other reasons, nowadays a putative non-barionic dark matter 
component permeating  the Universe is tacitly assumed.

Now at a quite different scale, spiral galaxies are known to have flat rotation curves, meaning that 
the graph of the velocity squared of test particles (stars, HII clouds,...) displays a {\it plateau} 
at about $5$kpc away from  galactic center extending as far as many tens of kiloparsecs. This results
is at odds with Newtonian prediction that at large distances the graph should fall off as $1/r$ --
unless we are witnessing here another manifestation of the same putative unseen cosmological material.
The manner dark matter clumps in the  galaxy is evaluated by recalling that a flat rotation curve 
must be generated by a logarithmic Newtonian  potential. Inserting this potential into Poisson's 
equation gives a density profile of the dark matter component  that falls off  as $1/r^2$. Adding up 
this component to the visible part accounts for $\Omega_{\mbox{Halo}} \approx 0.1$. Similar dynamical
 methods applied to the motion of clusters  due to their gravitational field at the scale of $20$ Mpc,
 yields $\Omega_{20} \approx 0.2$. Clearly this is a long way away from  $\Omega=1$ required from 
cosmological considerations. Therefore, consistency of the dark matter scenario requires a smoothly 
distributed component at scales larger than $20$ Mpc $\Omega_{\mbox{smooth}} \approx 0.8$ \cite{kolb}.

\subsection*{A must?}

The {\it raison d'\^{e}tre} of postulating a smoothly distributed and undetectable 
dark matter
component is just to yield the right bookkeeping for $\Omega=1$. If this were not enough, 
a pitfall awaits the dark matter scenario at the galactic scale. Tully and Fisher \cite{tully} 
discovered the empirical law  bearing their  name that relates the luminosity $L$ of a spiral galaxy
to the velocity at the  {\it  plateau}:
\begin{equation}               
V = 220 \mbox{km/s} (L/L_*)^{1/4} \, ,
\end{equation}
where  $L_*$ is a constant corresponding to the typical  luminosity of a galaxy \cite{peebles}. Now,
sources that contribute mostly to the luminosity in the frequency bandwidth  where the law is stated
correspond to  white dwarfs which, in turn, are mainly located  in the galactic disk. This brings 
about a conundrum \cite{bek} because if dark matter in the halo is to be blamed for the flat rotation
 curves, then a very fine tuning between disk and halo parameters would have to be called for, which
is hard to explain and even harder to implement. As a matter of fact, adjusting halo and disk 
parameters yields unavoidably to a ``bump'' in all rotation curves just before the {\it plateau} is
 reached, which is seldom observed \cite{bek}. The dark matter scenario becomes more intricate when 
one comes to the  question of its very nature  (massive  neutrinos, WIMPS,...), because all the 
candidates are of very hard direct detection.

A radically different approach would be to say that there is no considerable amount of dark matter 
permeating the Universe,  what we are rather witnessing in the spiral galaxies is the  breakdown of 
General Relativity  (at a given scale). Since galactic dynamics involves  weak gravitational 
fields and  non-relativistic motion, this clearly entails a modification of Newton's law too. Indeed, 
this was the step taken by Milgrom's \cite{milgrom} who put forward an explicit modification of the  
Newtonian dynamics that takes place when the Newtonian acceleration is of the order of  $a_0= 2 
\times 10^{-8} \mbox{cm/s}^{2}$ or smaller. He introduces a distinction between  the Newtonian  
gravitational field $\vec{g}_N$  and  the actual acceleration a test particle is subjected to, 
$\vec{g}$. In his proposal {\it Modified Newtonian Dynamics} (MOND, for brief) these accelerations
are related through:
\begin{equation}
\mu (\frac{g}{a_{0}})\vec{g} = \vec{g}_{N} ,
\end{equation}   where $\mu$ is a function satisfying,
\begin{equation}
\left\{
\begin{array}{ll}
       \mu (x) \rightarrow   x & \mbox{if $x<< 1$}  \\
       \mu (x) \rightarrow   1 & \mbox{if $x>> 1$  (Newtonian limit)} \, .
\end{array}
\right.
\end{equation}

MOND can be shown to \cite{bek}
\begin{itemize}
\item reproduce the flat rotation curves.
\item be consistent with Tully-Fisher's law. 
\item satisfy the weak equivalence principle but not the strong one. 
\end{itemize}

Extending this guideline into the relativistic domain clearly entails framing a new 
covariant theory. It would be a wise step to take General Relativity as a building block for such a 
theory because precision tests in the solar system seem to confirm General Relativity to a very 
high extend.  MOND suggests one to demand such a theory to comply to the weak 
equivalence  principle but not to the strong one. Furthermore, stability considerations 
requires a positive energy flux. A further imposition is that  causality must not be violated
at any rate. A first candidate, AQUAL  ({\it Aquadratic Lagrangian Theory}) \cite{review} was proposed
in the early eighties but was soon shown not to be a viable candidate because it was plagued with
superluminal propagation. The most promising candidate nowadays was proposed by Bekenstein and Milgrom  
\cite{review} an was baptized as {\it Phase Coupled Gravity} (PCG, for brief). In this theory, in addition to the 
metric tensor, $g_{\mu  \nu}^{*}$ the gravitational interactions are mediated by a complex scalar 
$\chi$  field. The corresponding action is 
\begin{equation} 
S_{\chi} = -\frac{1}{2}\int \left( g^{\alpha \beta} \chi_{, \alpha} \chi_{,\beta} + 
V(\chi^{*}\chi)\right) \sqrt{-g} d^{4}x \, ,
\end{equation} 
where  $V(x)$ represents the scalar field self-interaction. One expresses 
this action in a more convenient form by decomposing $\chi$ in terms of its amplitude $q$ and 
phase $\psi$:
\begin{equation}
S_{q, \phi}= - \frac{1}{2} \int \left( q_{,\alpha} q^{,\alpha} + q^{2}\phi_{,\alpha}\phi^{,\alpha} +
 V(q^{2})\right)\sqrt{-g}d^{4}x \, .
\end{equation}  
PCG is defined via the composition of this action with Einstein-Hilbert's and the
matter action which is defined through the replacement rule  $L_m \rightarrow
e^\phi L_m$. Put into words, matter couples only to the phase of the complex scalar field.
Clearly, predictions depend upon the choice of the potential $V(q^2)$. Minimal PCG ($V(x)=0$)
and the sextic  potential were show to lead to instabilities \cite{sanders1}. 

There are two alternative and equivalent representations of a scalar tensor theory. The first,
written  in i. Einstein's frame ($g_{\mu  \nu}^{*}$) where the scalar field interacts directly
with matter and test  particles do not follow geodetic lines; ii. the physical frame in which
the scalar field is absorbed  by the metric tensor via a conformal transformation
\begin{equation} 
g_{\mu  \nu} = g_{\mu  \nu}^{*}e^{ -\eta \phi} \, ,
\end{equation}
 where $\eta$ is some parameter.  In the physical frame the PCG action takes the
form:
\begin{equation}
 S_{f} = \frac{1}{16\pi G_0} \int \sqrt{-g}d^4x  e^{\eta \phi}\left[R -
q_{;\alpha}q^{;\alpha} - (q^{2} - \frac{3}{2} \eta^{2} )\phi_{;\alpha}\phi^{;\alpha}-  e^{\eta
\phi} V(q^{2}) \right]  + S_m.
\label{acao}
\end{equation} 
Here, $R$ is the scalar curvature and $G_0$, Newton's constant. Inspection of this equation reveals
that PCG corresponds to a  Brans-Dicke  theory with variable $\omega_{BD} = q^2 -3 \eta/2$. In order
to grapple with cosmological issues, a definite choice of the potential is needed. As we said, 
minimal PCG was discarded from stability grounds and the next simple candidate is a quadratic 
potential $V(q^2) = A q^{2} + B$. In order to reproduce the observed (flat) rotation curves in a 
flat Universe, Sanders obtained as the best fit for these parameters $A = 4,0 \times 10^4$, $B = 6.7$
and $\eta = 10^{-7}$ \cite{sanders2}. In this paper we shall study the very early Universe
in the framework of this particular model and, in particular, study the evolution of primordial
density fluctuations. 

\section{PCG Early Universe.}

 Sanders \cite{sanders2} obtained the evolution of FRW models in the framework of PCG  
solving numerically the differential equations for $a(t)$, $q(t)$  and $\phi (t)$. He obtained 
that at the very early universe (equation of state $p = - \rho$) the last two quantities are  
nearly constant. Inspired by his results, we took the {\it ansatz} $q(t) = \mbox{const.}$
and solved the flat model equations. Consistently, we obtained a large and slowly varying field 
$\phi(t)$. Then we studied  the fate of primordial density  perturbations in the early Universe. 
In contrast to the standard inflationary scenario, density perturbations are shown do grow during 
PCG inflation.

Variations of the action (\ref{acao}) with respect to the metric and both scalar fields yield the
following equations: 
\begin{itemize}
\item{metric variations}
\begin{eqnarray}          
 8 \pi G_0  T^{\alpha \beta} = (R^{\alpha \beta} - \frac{1}{2} R g^{\alpha \beta}) +
\frac{1}{2}q_{;\mu} q^{;\mu} g^{\alpha \beta}  + \nonumber \\ ( q^{2} - 
3/2 \eta ^{2}) (\frac{1}{2}\phi_{;\mu}\phi^{;\mu} g^{\alpha \beta} - \phi^{;\alpha} \phi^{;\beta}) 
- q^{;\alpha} q^{;\beta} +  \frac{1}{2}e^{\eta \phi} V(q^2) g^{\alpha \beta}   + \nonumber \\
\eta (\eta \phi_{;\mu} \phi^{; \mu}
+ \phi^{;\mu}_{; \mu}) g^{\alpha \beta} - \eta(\eta \phi^{;\beta} \phi^{;\alpha}
+ \phi^{;\alpha ;\beta})                                           
\label{g}
\end{eqnarray}
where $T^{\alpha \beta}$ is the energy momentum tensor of matter.

\item{$\phi$ variations}
\begin{eqnarray}   
\eta R - \eta q_{;\alpha} q^{;\alpha} + \eta (q^{2} - \frac{3 \eta^2}{2}) 
\phi_{;\alpha}\phi^{;\alpha} - 2V(q^2) \eta e^{\eta\phi} \nonumber \\
+ 4q q_{;\alpha} \phi_{;\beta} g^{\alpha \beta} + 2 (q^{2}- \frac{3 \eta^{2}}{2}) \Box\phi = 0
\label{phi} 
\end{eqnarray}                                                                        

\item{$q$ variations}:
\begin{equation}   
 e^{- \eta \phi}\left[e^{\eta\phi} q^{; \alpha}\right]_{; \alpha} - q \phi^{; \alpha} 
\phi_{; \alpha} - V^{'}(q^2) q e^{\eta \phi} = 0   \, .  
\label{q}
\end{equation}
\end{itemize}

A handy equation follows after combining  eq. (\ref{phi})  with the trace of eq. (\ref{g})
\begin{equation}
  4\pi G_0 T = (q^{2} \phi_{;\alpha} \phi^{;\alpha} + 
\frac{2q q_{;\alpha} \phi^{;\alpha}}{\eta} + 
 \frac{q^{2} \Box\phi}{\eta}) e^{\eta\phi} 
\label{trace}
\end{equation}
where  $T$ is the trace of the energy-momentum  tensor.

For a homogeneous and isotropic model the cosmological equations are enormously simplified. Indeed
the $00$ component of eq. (\ref{g}) reduces to
\begin{equation}  
\frac{ 8\pi G_0 }{3} \rho e^{-\eta \phi} = (\frac{\dot{a}}{a} + \frac{\eta\dot{\phi}}{2})^{2}
- \frac{\dot{q}^{2}}{6} - \frac{1}{6} e^{\eta\phi} V(q^2)  - \frac{1}{6} q^{2} \dot{\phi}^{2} \, ,
\label{gt}                                                      
\end{equation}
Similarly, eqs.  (\ref{phi}) and (\ref{q}),) collapse  respectively to 
\begin{eqnarray}  
\eta R + \eta \dot{q}^2 - \eta (q^2 - \frac{3}{2} \eta^2 ) \dot{\phi}^2 - 
2 V(q^2) \eta e^{\eta\phi} - 4 q \dot{\phi} \dot{q} - \nonumber \\
2 (q^2 - \frac{3}{2} \eta^2 ) \ddot{\phi} - 
6 \frac{\dot{a}}{a} (q^2 - \frac{3}{2} \eta^2 )\dot{\phi} = 0
\label{phit}
\end{eqnarray}
and
\begin{equation}
  \ddot{q} + 3\frac{\dot{a}}{a}\dot{q} + \eta\dot{\phi}\dot{q} - q \dot{\phi}^{2} = 
- q V'(q^2) e^{\eta\phi} \, .
\label{qt}                                                                     
\end{equation}

Finally, eq. (\ref{trace}) goes into:
\begin{equation}
  4\pi G_0 T = e^{\eta\phi} (-q^{2} \dot{\phi}^{2} - \frac{2}{\eta} q \dot{q} \dot{\phi}   - 
\frac{q^{2} \ddot{\phi}}{\eta} 
         - \frac{3}{\eta} q^{2} \frac{\dot{a}}{a} \dot{\phi}) 
\label{tracet}                                       
\end{equation}

Taking as the background solution $q_{0} = $ const, it follows from eq. (\ref{qt}) that:
\begin{equation}
\dot{\phi}_0^2 = A e^{\eta \phi_{0}} \, . 
\label{seila}
\end{equation}    
Integration is trivial, 
\begin{equation}
\phi_{0} = \frac{-2}{\eta} \ln (-\frac{\eta \sqrt{A}t}{2}   + K)   
\label{phi0}
\end{equation}
where $K$ is an integration constant. As  anticipated, $\phi(t)$ is a slowly varying function in the 
early Universe. Having obtained the evolution of the background fields in the early Universe, 
we pursue the analysis of the behavior of perturbations:
\begin{equation}
\left\{\begin{array}{ll}
\rho &= \rho_{0} + \rho_{1} \\
p &= p_{0} + p_{1} \\
\phi &= \phi_{0} + \phi_{1} \\
q &= q_{0} + q_{1} \\
 g_{\alpha  \beta} &= a^{2}\delta_{\alpha  \beta}  +  h_{\alpha \beta}\\
\end{array}                                  
\right.
\end{equation}

The evolution of those perturbations are obtained through the linearization of eq's. (\ref{g}), 
(\ref{q}) and (\ref{trace}). After some tedious but straightforward algebra we obtained the
\begin{itemize}
\item metric perturbations ($00$ component)
\begin{eqnarray}
8\pi G_0 (\rho_1 - \rho_0 \eta \phi_1) & = e^{\eta\phi_0}
[ R^{*}_{00} + \frac{1}{2} R^* - 
 \dot{q}_0 \dot{q}_1 - (q_0^2 - \frac{3}{2}\eta^2 )(\dot{\phi}_0 \dot{\phi}_1) - 
\nonumber \\ & q_{0} q_{1} \dot{\phi}_0^2 - 
 A q_{0} q_{1}e^{\eta \phi_0}  - e^{\eta \phi_0}\frac{\eta}{2} \phi_{1} (A q_{0}^{2} + B) - \nonumber \\
& \eta \Box q_1 - \eta \ddot{\phi}_1] \,
\label{gp}
\end{eqnarray} 
where $R^*$ e $R^*_{00}$ are the perturbations of the scalar curvature the 
Ricci's tensor $oo$ component.

\item q-perturbations
\begin{equation} 
-\eta \dot{\phi}_0 \dot{q}_1 -\eta \dot{\phi}_1 \dot{q}_0 + 
 q_{1 ;\alpha}^{;\alpha}  +  q_{1} \dot{\phi}_0^2 + 2 \dot{\phi}_0 \dot{\phi}_1 q_0 +
A e^{\eta\phi_{0}} ( q_1 + \eta q_0 \phi_{1}) = 0    
\label{qp}
\end{equation} 

\item  $\phi$ perturbations 
\begin{eqnarray}
4 \pi G_0 e^{- \eta \phi_{0}} (T_{1} - \eta \phi_{1} T_{0}) = & - 2q_{0}^{2}\dot{\phi}_0\dot{\phi}_1
 - 2q_0 q_1 \dot{\phi}_0^2 - \frac{2}{\eta} q_{0}(\dot{q}_0 \dot{\phi}_1 + \dot{q}_1 \dot{\phi}_0)  \nonumber \\
 & - \frac{2}{\eta} q_{1}\dot{q}_0\dot{\phi}_0 +  
\frac{1}{\eta} q_{0}^{2} \Box \phi_1  +  
\frac{2}{\eta} q_{0} q_{1} \Box\phi_{0} \, .
\label{phip}
\end{eqnarray}
\end{itemize}

Assuming a period of inflationary evolution where $a = e^{\Lambda t}$ (and, correspondingly, an
equation of state  $p = - \rho$) and metric perturbations of the form $h_{ij}= 1/3 h \delta_{ij}$,
 it follows from the perturbations of the $oi$ component of  eq.(\ref{g}) that :
\begin{equation}
 \dot{h} - 2 h\Lambda = - 3 e^{2 \Lambda t} (\dot{\phi}_1 - \Lambda \phi_{1})  , 
\label{lapse}
\end{equation}   
where $h  = h_{1 1} + h_{2 2} + h_{3 3}$.

We are already  in position of solving the equations for the perturbations. Specializing for the case
where the background fields are exactly given by eq (\ref{phi0}) and $q_0 =$ const and further 
defining a new time variable $\tau = - \frac{\eta t}{2} + \frac{K}{\sqrt{A}}$ we obtain  
from eq. (\ref{phip}) in the vanishing wave number limit:
\begin{itemize}
\item  the evolution of $\phi_1$
\begin{equation}
\phi_1^{''} + (\chi_1 - \frac{5 }{\tau} )\phi_{1}^{'} + \phi_{1} (\chi_{2} + 
\frac{\chi_{3}}{\tau} + \frac{\chi_{4}}{\tau^2} ) + 
\frac{\chi_5 q_1^{'}}{\tau}  +q_1 (\frac{\chi_6}{\tau} + \frac{\chi_7}{\tau ^2})  = 0 \, ,
\label{phi1}
\end{equation}
from eq.(\ref{qp}) 
\item the evolution of $q_1$
\begin{equation}
 q_{1}^{''} + q_{1}^{'}(\alpha - \frac{2}{\tau}) + 
\frac{\chi_{8}\phi_{1}^{'}}{\tau} 
 + \frac{\chi_{8} \phi_{1}}{\tau^{2}} = 0 \, ,    
\label{q1}
\end{equation}

and, finally from eq. (\ref{gp})
\item  the evolution for $\rho_1$
\begin{eqnarray}
8 \pi G_0 (\rho_1 - \eta \rho_{0} \phi_{1} ) & =   \left\{  \dot{\phi}_1 \left[ - ( q_0^2 - 
\frac{3}{2} \eta^{2})   \frac{2 \sqrt{A}}{2 K - \eta \sqrt{A} t} - 
3 \eta^{2}  \frac{\sqrt{A}}{2 K- \eta \sqrt{A}t } \right] \right.     
\nonumber \\
& +    \phi_1 \left[ 3 \Lambda^2 \eta - 2 \eta \frac{A q_0^2 + B}{(2 K - \eta \sqrt{A}t)^2} 
 +  3  \eta^{2} \Lambda  \frac{\sqrt{A}}{2 K- \eta \sqrt{A} t} \right] 
\nonumber \\ 
 & + q_{1}\left[ \left. \frac{-8 q_0 A}{(- \eta \sqrt{A} t + 2 K)^2 } \right] \right\} 
\frac{4}{(+2K- \eta \sqrt{A} t)^2} \, .      
\label{rho1}
\end{eqnarray}
\end{itemize}
In the above equations, primes represent derivatives with respect to  $\tau$ and the $\chi$'s
are constants displayed in the table beneath:
\begin{equation}
\begin{array}{||r|l||}
\hline
\multicolumn{2}{|c|}{\mbox{Cosmological Parameters}}\\ \hline\hline
\chi_1 &  -\frac{6\Lambda}{\eta}   \\ \hline
\chi_2 &  -\frac{6 \Lambda^2 \eta^2}{q_0^2}   \\ \hline
\chi_3 &  \frac{3\eta \Lambda}{2} -\frac{3 \eta^3 \Lambda}{q_0^2}   \\ \hline
\chi_4 &  \frac{\eta^2}{A q_0^2}(Aq_0^2 + B)   \\ \hline
\chi_5 &  -\frac{\eta}{q_0}   \\ \hline
\chi_6 &  \frac{6 \Lambda}{q_0}   \\ \hline
\chi_7 &  \frac{7 \eta}{q_0}  \\ \hline
\chi_8 &  \frac{4 q_0}{\eta}   \\ \hline
\alpha &  -\frac{6 \Lambda}{\eta}   \\ \hline
\end{array}
\label{tabela}
\end{equation}

Integrating eq. (\ref{q1}), it follows that 
\begin{equation}
\phi_1 = \frac{1}{\chi_8 \tau}\left\{q_1^{'} \tau^2  + \alpha \tau^2 q_1 -
4  q_1 \tau +
\psi \right\}\, ,
\label{integral}
\end{equation}
where $\psi = \int q_1 (4 - 2\alpha\tau) d\tau$.
 
Putting together this result with eq. (\ref{phi1}), yields the differential equation
\begin{eqnarray}
& q^{'''}_1 \tau + q_1^{''}\left[-7 + \tau (\alpha  + \chi_1 ) \right]
\nonumber \\
& + q_1^{'}\left[ 
\frac{1}{\tau}(  19 +\chi_4 +\chi_8 \chi_5) + ( \chi_3 - 3\chi_1 )
 + \tau(\alpha\chi_1 - 5 \alpha + \chi_2)\right] 
\nonumber \\
& q_1 \left[\frac{1}{\tau^2}(-28 - 4\chi_4 + \chi_7\chi_8)
 + \frac{1}{\tau}(7 \alpha + 4\chi_1 + \alpha\chi_4 - 4\chi_3 + \chi_8 \chi_6) + (-\alpha \chi_1 + 
\alpha \chi_3 - 4 \chi_2)\right]\nonumber \\
& + \psi\left[ \frac{1}{\tau^3}(7 + \chi_4) + \frac{1}{\tau^2}(\chi_3 - \chi_1) + \frac{1}{\tau}(
\chi_2)\right] = 0 \, .
\label{colossal}
\end{eqnarray}
In the late time limit $\left| \tau \right| >> 1$, this hyper colossal equation boils down to
\begin{equation}
q_1^{'''} + q_1^{''}( \chi_1 + \alpha) + q_1^{'}(\alpha \chi_1 + \chi_2) +
 q_1 \alpha \chi_2 = 0\, .
\label{lateq}
\end{equation}
Inserting $q_{1} = e^{\frac{6 \Lambda Q \tau}{\eta}}$ into this equation an algebraic equation for 
$Q$ follows 
\begin{equation}
Q(Q - 1)^2  + \frac{\eta^4}{6q_0^2}(1 - Q) = 0 \, , 
\label{`Q'}
\end{equation} 
whose solutions for $Q$ are
\begin{equation}
Q = \left\{
\begin{array}{ll} 
1 \nonumber \\
 \frac{1}{2} \pm \frac{1}{2} \sqrt{1 + \frac{4\eta^4}{6q_0^2}}\, .
\end{array} 
\right.
\end{equation}
Assuming $\eta^4/q_0^2 << 1$, the negative root is approximately :
\begin{equation}
Q= - \frac{\eta^4}{6 q_0^2}
\label{Q}
\end{equation}
This result allows us to obtain $\phi_{1}$, through eq. (\ref{phi1}) 
\begin{equation} 
\phi_1 = \left( \frac{1}{\chi_8}(\frac{-\eta^3 \Lambda}{q_0^2} \tau + \alpha \tau - 4) + 
\frac{2 \alpha q_0^2}{ \eta^3 \Lambda}\right)
 e^{  \frac{-\eta^3 \Lambda}{q_0^2} \tau}
\label{latephi}
\end{equation}

Our concern in here is the  asymptotic behavior of $\rho_1$. This can be obtained  combining the 
the late time regime limit of eq. (\ref{rho1}) with eqs. (\ref{lateq}) and (\ref{latephi})
\begin{equation}
8\pi G_0 \rho_1 \approx   
\frac{e^{P \tau}}{\tau^2} \frac{\eta ( P + \alpha)}{\chi_8 A }\left\{\frac{\eta P q_0^2}{2} + 
3\Lambda^2 +  8 \pi G_0 \rho_0 )  \right\} \, ,
\label{final}
\end{equation}
where $P= - \eta^3 \Lambda q_0^{-2}$ 
This  exponential-like form of density perturbations should be contrasted  with the 
general relativistic prediction that they remain strictly constant during inflation \cite{kolb}. 

\section{Concluding Remarks}

In the standard cosmological scenario density perturbations are frozen during  the radiation
dominated era, and are allowed to grow only  after the decoupling between matter and radiation 
has taken place \cite{wein}. This might leave a very tight time-schedule for the contrast density to 
grow from $10^{-5}$ at $z \approx 1400$ to unity at a red-shift  of order 1. Likewise, in PCG 
the density perturbations grown during inflation are frozen during the radiation 
dominated era. During this era and onwards the PCG cosmological evolution is very much 
similar to that one of the standard model \cite{sanders2}. Consequently, many predictions of PCG 
cosmology are similar to those of the standard model. One of the main differences lies in that the 
exponential-like grow of $\rho_1$ accumulated during inflation will be carried over to the time of 
decoupling when the perturbations are finally allowed to grow. Therefore, PCG predicts a much larger 
density contrast, which could alleviate the problem of the tight time-schedule for the contrast
density  to enter into the non-linear regime.

One of the crucial checks of the standard cosmological scenario is the abundance of
light elements. How does  PCG predictions of light elements compare to the standard model?
As discussed by Sanders \cite{sanders2}, the version of PCG with a quadratic potential
leads to a somewhat  faster expansion of the Universe during  nucleosynthesis (of
about $6 \%$), causing an apparent  overproduction of primordial Helium (earlier freeze-out of
neutrons). Nevertheless, the same increase in the expansion rate leads to a reduction in 
the abundance of neutrons and the two  effects would nearly compensate. Thus, the Helium abundance 
would remain insensitive to the faster expansion and would be within the present observational limits. 
Unfortunately, the same is not true for the heavier elements, the abundance of heavier nuclei 
like $H^{2}$  and $Li^{7}$  would be changed drastically by a factor from $10$ to a $100$ 
\cite{sanders2}.  Furthermore, it must be mentioned that the present version of PCG is burdened by 
other problems, and seems to be in disagreement with  precision experiments in the solar system, 
in particular the precession of the of perihelion of Mercury \cite{beksan}. Other versions of the 
theory should be explored in order to come to grips with the observational data.

The bottom line of this paper is to show that viable theories of gravity can be constructed to 
explain many of the cosmological paradoxes. Furthermore, in contrast to the dark matter scenario, 
where unseen matter can be placed here and there at will to justify the discrepancies between 
predictions and observations and with no further consequences, these theories produce many definite 
predictions which can be checked against the observational data. This fact does turn  these 
theories very atractive.

\section*{Acknowledgments} C.E.M.B. wants to express his thankfulness to ??? for a master 
scholarship and M.S.is thankful to CNPq for partial financial support.

\end{document}